\documentclass[aps,prc,reprint,superscriptaddress,letter]{revtex4-1}
\usepackage{graphicx}
\usepackage{bm}
\usepackage{amsfonts}
\usepackage{amsmath}
\usepackage{amssymb}
\usepackage{dcolumn}
\usepackage[autolanguage]{numprint}
\usepackage{color}
\usepackage[table]{xcolor}
\usepackage{psfrag}

\begin{document}

\preprint{V. 0.5}

\title{Confirmation of the isomeric state in \textsuperscript{26}P}


\author{D. P\'erez-Loureiro}
\email[]{perezlou@nscl.msu.edu}
\affiliation{National Superconducting Cyclotron Laboratory, Michigan State University, East Lansing, Michigan 48824, USA}
\author{C. Wrede}
\email[]{wrede@nscl.msu.edu}
\affiliation{National Superconducting Cyclotron Laboratory, Michigan State University, East Lansing, Michigan 48824, USA}
\affiliation{Department of Physics and Astronomy,  Michigan State University, East Lansing, Michigan 48824, USA}
\author{M.~B.~Bennett}
\affiliation{National Superconducting Cyclotron Laboratory, Michigan State University, East Lansing, Michigan 48824, USA}
\affiliation{Department of Physics and Astronomy, Michigan State University, East Lansing, Michigan 48824, USA}
\author{S.~N.~Liddick}
\affiliation{National Superconducting Cyclotron Laboratory, Michigan State University, East Lansing, Michigan 48824, USA}
\affiliation{Department of Chemistry, Michigan State University, East Lansing, Michigan 48824, USA}
\author{A.~Bowe}
\affiliation{National Superconducting Cyclotron Laboratory, Michigan State University, East Lansing, Michigan 48824, USA}
\affiliation{Department of Physics and Astronomy, Michigan State University, East Lansing, Michigan 48824, USA}
\affiliation{Physics Department, Kalamazoo College, Kalamazoo, Michigan 49006, USA}
\author{B. A. Brown}
\affiliation{Department of Physics and Astronomy,  Michigan State University, East Lansing, Michigan 48824, USA}
\affiliation{National Superconducting Cyclotron Laboratory, Michigan State University, East Lansing, Michigan 48824, USA}
\author{A.~A.~Chen}
\affiliation{Department of Physics and Astronomy, McMaster University, Hamilton, Ontario L8S 4M1, Canada}
\author{K.~A.~Chipps}
\affiliation{Department of Physics, Colorado School of Mines, Golden, Colorado 08401, USA}
\affiliation{Physics Division, Oak Ridge National Laboratory, Oak Ridge, Tennessee 37831, USA}
\affiliation{Department of Physics and Astronomy, University of Tennessee, Knoxville, Tennessee 37996, USA}
\author{N.~Cooper}
\affiliation{Department of Physics and Wright Nuclear Structure Laboratory, Yale University, New Haven, Connecticut 06520, USA}
%
%
\author{E.~McNeice}
\affiliation{Department of Physics and Astronomy, McMaster University, Hamilton, Ontario L8S 4M1, Canada}
%
%
\author{F.~Naqvi}
\affiliation{Department of Physics and Wright Nuclear Structure Laboratory, Yale University, New Haven, Connecticut 06520, USA}
\author{R.~Ortez}
\affiliation{Department of Physics and Astronomy, Michigan State University, East Lansing, Michigan 48824, USA}
\affiliation{National Superconducting Cyclotron Laboratory, Michigan State University, East Lansing, Michigan 48824, USA}
\affiliation{Department of Physics, University of Washington, Seattle, Washington 98195, USA}
\author{S.~D.~Pain}
\affiliation{Physics Division, Oak Ridge National Laboratory, Oak Ridge, Tennessee 37831, USA}
\author{J.~Pereira}
\affiliation{National Superconducting Cyclotron Laboratory, Michigan State University, East Lansing, Michigan 48824, USA}
\affiliation{Joint Institute for Nuclear Astrophysics, Michigan State University, East Lansing, Michigan 48824, USA}
\author{C.~Prokop}
\affiliation{Department of Chemistry, Michigan State University, East Lansing, Michigan 48824, USA}
\affiliation{National Superconducting Cyclotron Laboratory, Michigan State University, East Lansing, Michigan 48824, USA}
%
%
\author{S.~J.~Quinn}
\affiliation{Department of Physics and Astronomy, Michigan State University, East Lansing, Michigan 48824, USA}
\affiliation{National Superconducting Cyclotron Laboratory, Michigan State University, East Lansing, Michigan 48824, USA}
\affiliation{Joint Institute for Nuclear Astrophysics, Michigan State University, East Lansing, Michigan 48824, USA}
\author{J.~Sakstrup}
\affiliation{Department of Physics and Astronomy, Michigan State University, East Lansing, Michigan 48824, USA}
\affiliation{National Superconducting Cyclotron Laboratory, Michigan State University, East Lansing, Michigan 48824, USA}
\author{M.~Santia}
\affiliation{Department of Physics and Astronomy, Michigan State University, East Lansing, Michigan 48824, USA}
\affiliation{National Superconducting Cyclotron Laboratory, Michigan State University, East Lansing, Michigan 48824, USA}
\author{S.~B.~Schwartz}
\affiliation{Department of Physics and Astronomy, Michigan State University, East Lansing, Michigan 48824, USA}
\affiliation{National Superconducting Cyclotron Laboratory, Michigan State University, East Lansing, Michigan 48824, USA}
\affiliation{Geology and Physics Department, University of Southern Indiana, Evansville, Indiana 47712, USA}
\author{S.~Shanab}
\affiliation{Department of Physics and Astronomy, Michigan State University, East Lansing, Michigan 48824, USA}
\affiliation{National Superconducting Cyclotron Laboratory, Michigan State University, East Lansing, Michigan 48824, USA}
\author{A.~Simon}
\affiliation{National Superconducting Cyclotron Laboratory, Michigan State University, East Lansing, Michigan 48824, USA}
\affiliation{Department of Physics and Joint Institute for Nuclear Astrophysics, University of Notre Dame, Notre Dame, Indiana 46556, USA}
\author{A.~Spyrou}
\affiliation{Department of Physics and Astronomy, Michigan State University, East Lansing, Michigan 48824, USA}
\affiliation{National Superconducting Cyclotron Laboratory, Michigan State University, East Lansing, Michigan 48824, USA}
\affiliation{Joint Institute for Nuclear Astrophysics, Michigan State University, East Lansing, Michigan 48824, USA}
\author{E.~Thiagalingam}
\affiliation{Department of Physics and Astronomy, McMaster University, Hamilton, Ontario L8S 4M1, Canada}


\date{\today}

\begin{abstract}
We report the independent experimental confirmation of an isomeric state in  the proton drip-line nucleus \textsuperscript{26}P. 
The $\gamma$-ray energy and half-life determined are $164.4\pm0.3\,(\text{sys})\pm0.2\,(\text{stat})$~keV and $104\pm 14$~ns, respectively, which are in agreement with the previously reported values. These values are used to set a  semi-empirical limit on the proton separation energy of \textsuperscript{26}P, with the conclusion that it can be bound or unbound.  
\end{abstract}


\maketitle

\section{Introduction\label{sec:intro}}
\textsuperscript{26}P is a very proton-rich nucleus close to the proton drip-line that ${\rm \beta}$ decays ($t_{1/2}=43.7\pm0.6$ ms)\cite{Thomas2004}. The ground state was discovered in 1983 by Cable \emph{et al.} \cite{Cable1983,Thoennessen2012} and the tentative spin and parity is $J^{\pi}=(3^+)$ \cite{BASUNIA20161}. Its predicted low proton separation energies ($143\pm200$~keV \cite{AME2012}, $0\pm90$~keV\cite{Thomas2004}), together with the narrow momentum distribution and enhanced cross section, both observed in proton-knockout reactions \cite{Navin1998}, as well as a significant mirror asymmetry in $\mathrm{beta}$ decay \cite{DPL2016}, give experimental evidence for the existence of a proton halo \cite{Tanihata2013,Brown1996,Ren1996,Gupta2002,Liang2009}. It is even possible that \textsuperscript{26}P is unbound to proton emission, as various mass models predict \cite{Moller2016,Goriely2014,Koura2005}, but $\mathrm{\beta}$ decays instead due to the Coulomb barrier. In a recent experiment \citet{Nishimura2014} reported the observation of an isomeric state with $J^\pi=1^+$ in \textsuperscript{26}P. This state is the mirror analog  of the low-lying isomer of \textsuperscript{26}Na which has an excitation energy of $82.5\pm0.5$ keV \cite{Dufour1987}. The reported excitation energy and half-life of the \textsuperscript{26}P state were $164.4\pm0.1$~keV and $120\pm9$~ns, respectively \cite{Nishimura2014}. In this paper we report confirmation of this isomeric state in an independent experiment \cite{Bennett2013,Schwartz2015,DPL2016} using a different production mechanism and a different setup at a different facility. 
\section{Experiment\label{sec:exp}}
The experiment was carried out at the National Superconducting Cyclotron Laboratory (NSCL) at Michigan State University. A primary beam of \textsuperscript{36}Ar with an intensity of 75~pnA was accelerated by the Coupled Cyclotron Facility to an energy of 150~MeV/u and impinged upon a 1.5~g/cm\textsuperscript{2} Be target. The \textsuperscript{26}P ions produced via nuclear fragmentation were separated in flight from other reaction products by the A1900 fragment separator \cite{Morrissey200390}. A 120~mg/cm\textsuperscript{2} wedge-shaped Al degrader was placed at the dispersive plane of the spectrometer to separate the incoming fragmentation residues according to their atomic charge and  thus enhance the beam purity.
 The secondary beam was further purified by means of the Radio Frequency Fragment Separator \cite{Bazin2009314} and implanted into a planar germanium double-sided strip detector (GeDSSD) \cite{Larson201359}. The $\mathrm{\gamma}$ rays emitted in coincidence with the implantation signals were detected by the high-purity segmented germanium array (SeGA) \cite{Mueller2001492}. More details about the experimental set up can be found in Refs. \cite{Bennett2013,Schwartz2015,DPL2016}. 

The isotopic identification of the secondary beam particles was accomplished by measuring the energy loss and time-of-flight of the incoming nuclei ($\Delta E$-ToF method). $\Delta E$ signals were provided by a pair of silicon detectors placed 1~m upstream from the GeDSSD. The ToF was measured between a 13.1-mg/cm\textsuperscript{2}-thick plastic scintillator located 25~m upstream, at the focal plane of the A1900, and one of the silicon detectors \cite{DPL2016}.

The data were collected event-by-event with the NSCL digital data acquisition system \cite{Prokop2014}. Each channel provided its own time-stamp signal, which made it possible to set coincidence gates between different detectors. Implantation events were selected  by requiring coincident signals between the silicon detectors and the GeDSSD.  

\section{Analysis, Results, and  Discussion\label{sec:results}}

\begin{figure}[t]
\centering
 \includegraphics[width=0.49\textwidth]{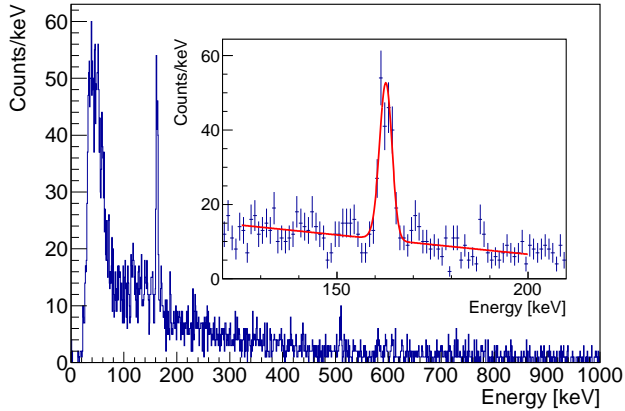}%
 \caption{(color online)  $\gamma$-ray energy spectrum associated with \textsuperscript{26}P implantations with a time gate of 2~$\mu$s between the silicon detector and the SeGA signals (blue online). The inset shows a magnification of the peak region including statistical error bars and the corresponding fit function is represented by the smooth solid line (red online). \label{fig:spectrum}}
 \end{figure}

\subsection{Energy}
A two-dimensional software gate was applied in the $\Delta E$-ToF identification matrix of the implanted ions to select the \textsuperscript{26}P nuclei \cite{DPL2016}. This made it possible to isolate the $\mathrm{\gamma}$ rays emitted in coincidence with \textsuperscript{26}P implantations. The 16 spectra obtained from the individual elements of SeGA were then added together after they were gain matched. The resulting spectrum was then calibrated in energy \cite{DPL2016}. Figure \ref{fig:spectrum} shows the $\rm \gamma$-ray spectrum corresponding to \textsuperscript{26}P implantations within a 2-$\rm \mu$s window. The spectrum shows a clear peak at $164.4\pm0.3\,(\text{sys})\pm0.2\,(\text{stat})$~keV. The peak energy was obtained by fitting the photopeak with an exponentially modified Gaussian (EMG) function summed with a linear function to model the local background.
\subsection{Half-life\label{sec:half-life}}
\begin{figure}[t]
\centering
\includegraphics[width=0.49\textwidth]{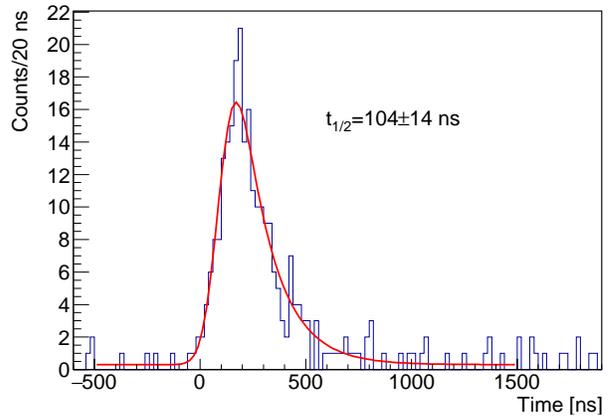}%
\caption{(color online) Distribution of time differences between $\gamma$-ray signals in SeGA gated on the 164-keV peak and \textsuperscript{26}P-ion signals in the silicon detector (blue online). The smooth solid line (red online) corresponds to the EMG fit function discussed in Sec. \ref{sec:half-life}. \label{fig:time}}
\end{figure}

The half-life of the state was determined from a fit of  the time distribution of the $\rm \gamma$-ray signals in SeGA with respect to the \textsuperscript{26}P ion signals in the silicon detector included within a gate centered at the energy of the peak and 10~keV wide. Figure~\ref{fig:time} shows the distribution of the time difference between the silicon detector, which provided the start time for the decay gate, and the SeGA germanium array signals, provided an implant signal was registered in the GeDSSD. Two different fit functions were employed: the first one is an EMG summed with a constant background, but with a negative decay parameter $rm \tau$. To verify the width and centroid of this fit, we checked that the results were consistent with the width and centroid obtained by fitting a Gaussian peak shape to the time spectrum of prompt $\gamma$ rays in the energy spectrum. The other one is an exponential decay added to a constant background. The range of this latter fit function was between the maximum of the time distribution and 1.5 $\mu$s. Both fits were performed using the maximum likelihood method. This method made it possible to account for low statistics and empty bins in the background region. The results of both fits were consistent within uncertainties. The measured half-life using the EMG fit is $t_{1/2}=104\pm14$~ns,  which is in very good agreement with $t_{1/2}=120\pm9$~ns reported by \citet{Nishimura2014}, but slightly lower. 

\subsection{Isomeric ratio}
The isomeric ratio $R$ is defined as the probability that, if a \textsuperscript{26}P nucleus is produced in the reaction, it is produced in an isomeric state. It is given by the following equation \cite{Pfutzner2002}:
\begin{equation}
R=\frac{Y}{N_{\text{imp}}FG}, 
\label{eq:ratio}
 \end{equation}
where $Y$ is the observed isomer yield at the decay station, $N_{\text{imp}}$ is the number of implanted \textsuperscript{26}P ions, $F$ and $G$ are correction factors for in-flight decay losses and nuclear reactions in the GeDSSD that destroy a fraction of the produced isomers, respectively. $Y$ is calculated as
\begin{equation}
Y = \frac{N_\gamma(1+\alpha_{\text{tot}})}{\varepsilon}, 
\label{eq:yield}
 \end{equation}
where $N_\gamma$ is the number of counts in the 164-keV peak, $\alpha_{tot}$ is the total conversion coefficient for this transition, and $\varepsilon$ is the $\gamma$-ray detection efficiency. $N_\gamma=175\pm17$ was obtained from the area below the photopeak fit. The efficiency $\varepsilon=(13\pm2)\%$ was determined using the calibration of Ref. \cite{DPL2016} and $\alpha_{tot}=0.0188\pm0.0003$ was estimated using the online calculator {\sc bricc} \cite{Kibedi2008}, under the assumption that the 164-keV transition has an $E2$ multipolarity \cite{Nishimura2014}.

 The correction factor $F$ is calculated as
\begin{equation}
F = \exp{\left[-\frac{1}{\tau} \left (\frac{\text{ToF}_1}{\gamma_1} + \frac{\text{ToF}_2}{\gamma_2} + \frac{\text{ToF}_3}{\gamma_3}\right )\right ]}
\label{eq:f_factor}
\end{equation}
where $\tau=150\pm 20$~ns is the mean lifetime of the state and $\text{ToF}_{1(2)}$ and $\gamma_{1(2)}$ are the time-of-flight and Lorentz factors through the first (second) section of the A1900, respectively. $\text{ToF}_{3}$ and $\gamma_{3}$ correspond to the time of flight and  the Lorentz factor for the flight path between the focal plane of the A1900 and the decay station. $\text{ToF}_{i}$ and $\gamma_{i}$ were calculated using the {\sc lise++} code \cite{Tarasov2008}, taking into account the thicknesses of all the different  layers of matter traversed by the secondary beam. The value of this  correction factor is  $F=0.03\pm0.01$. $G$ was obtained by calculating with {\sc lise++} the survival probability after traversing 400 $\mu$m of germanium, which corresponds to the average implantation depth in the GeDSSD. The value of $G$ is $0.995\pm0.005$.  The isomeric ratio obtained after applying these corrections is $R=(14\pm10)\%$, which is much lower than the $97^{+3}_{-10}\%$ reported by \citet{Nishimura2014} using a $^{28}$Si beam impinging on a  polyethylene target to produce \textsuperscript{26}P. This difference in  the isomeric ratios  may be explained by the different reaction mechanisms used to produce  \textsuperscript{26}P.
Future reaction experiments with \textsuperscript{26}P secondary beams may now select either the ground state or the isomeric state by exploiting the different isomeric ratios obtained depending on the reaction mechanism used to produce the  radioactive beam.

It is also worth mentioning that previous experiments using \textsuperscript{26}P, like the one reported by \citet{Navin1998}, would have had this isomeric state in their beam. However, because of the production mechanism, the short half-life of the state, and the long pathlength between the production and reaction targets (70 m) \cite{Navin1998}, only 0.2\% of  the \textsuperscript{26}P nuclei impinging on the secondary target would correspond to the isomer in this case. Such a small amount would not affect significantly the results reported in Ref. \cite{Navin1998}. 
\subsection{Estimation of $^{\mathbf{26}}$P proton separation energy}xs
\begin{figure}[t]
\centering
\includegraphics[width=0.49\textwidth]{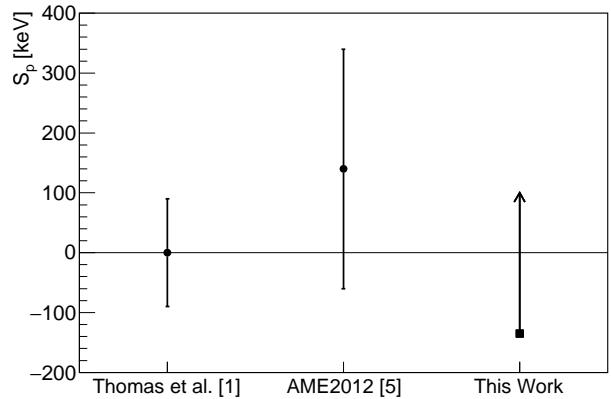}%
\caption{Comparison of semiempirical estimates of the  proton separation energy of \textsuperscript{26}P present in literature (circles) \cite{Thomas2004,AME2012}, with the lower limit obtained in this work (square).\label{fig:separation}}
\end{figure}

If the isomeric state was far above the proton separation energy of \textsuperscript{26}P, it would likely decay by emitting protons instead of gamma rays, as observed.
We can therefore use the  measured values of the energy and half-life of this isomeric state to set a semi-empirical limit on the proton separation energy of \textsuperscript{26}P, which is not known experimentally. We know from Ref. \cite{Nishimura2014} that the branching ratio for proton emission from this state is at most 13\%. The $\gamma$-ray  and  proton  partial widths  are therefore related as
\begin{equation}
\Gamma_p\leq \frac{13}{87} \Gamma_\gamma.
\label{eq:Gamma}
\end{equation}
The partial width for $\gamma$ rays obtained from our half-life result is $\Gamma_\gamma=4.39\pm0.59$~neV.

$\Gamma_p$  is related to the energy of resonant proton capture $E_r$ by the following equation \cite{Iliadis1997}:
\begin{equation}
\Gamma_p=\frac{2\hbar^2}{\mu R_n^2}P_\ell(E_r,R_n)C^2S\theta^2_{\text{sp}}.
\label{eq:Gamma_p}
\end{equation}
In this expression $R_n$ is the interaction radius [$1.25(1^{1/3}+25^{1/3})$~fm for this case], $\mu$ is the reduced mass of the system, $P_\ell$ is the barrier penetration factor, $C$ is an isospin Clebsch-Gordan coefficient, $S$ is the spectroscopic factor, and $\theta^2_{\text{sp}}$ is the single particle reduced width \cite{Iliadis1997,Barker1998}. The penetration factor may be calculated as $P_\ell(E_r,R_n)=kR_n/(F_\ell^2+G_\ell^2)$, where $k$ is the wave number and $F_\ell(G_\ell)$ is the regular (irregular) Coulomb wave function.

To set a limit on the proton separation energy of \textsuperscript{26}P with the obtained experimental results, we solved  Eq.~\eqref{eq:Gamma_p} for the kinetic energy ($E_r$), such that the proton emission width equals  the limit of the inequality in Eq.~\eqref{eq:Gamma}. The  values of the spectroscopic factors, $C^2S(3/2)=0.23\pm0.02$ for the 0d$_{3/2}$ shell, and $C^2S(5/2)=0.13\pm0.01$ for the 0d$_{5/2}$ shell,  were obtained from shell model calculations using the universal sd version B Hamiltonian \cite{Brown2006}, and the single particle widths were calculated using the parametrizations given in Refs. \cite{Iliadis1997,Barker1998}. The value obtained for the $^{25}\mathrm{Si+p}$  center-of-mass kinetic energy is therefore $E_r\leq300$~keV, where a single particle width of $\theta^2_{\text{sp}}=0.35\pm0.04$ was employed.

The kinetic energy ($E_r$), the excitation energy  ($E^*$), and the separation energy ($S_p$) are related as
\begin{equation}
E^*=E_r+S_p.
\label{eq:S_p}
\end{equation}
Thus, solving  Eq.~\eqref{eq:S_p} for $S_p$ using $E^*=164.4\pm0.4$ keV and the resonance energy calculated previously, the value obtained for the proton separation energy of \textsuperscript{26}P  is  $S_p\geq-135$~keV.  Figure \ref{fig:separation} shows a comparison of the lower limit obtained in this work with the two values for the proton separation energy of  \textsuperscript{26}P  in the literature. The first of these two literature values was deduced using  the prediction of the mass excess of \textsuperscript{26}P from systematic extrapolations given  in the atomic mass evaluation (AME) \cite{AME2012}. The second one was obtained by \citet{Thomas2004} using the Coulomb energy difference from \textsuperscript{26}Si  and the energy of the isobaric analog state using the semiempirical Isobaric Multiplet Mass Equation \cite{Antony1986}. We observe that our result is consistent with previous results, both compatible with a loosely bound (or unbound) valence proton in \textsuperscript{26}P. 

\section{Conclusions\label{sec:Conclusions}}

We have observed  a $164.4\pm0.3(\text{sys})\pm0.2(\text{stat})$~keV peak in the gamma-ray spectrum emitted in coincidence with  an implanted \textsuperscript{26}P secondary beam produced by $^{36}$Ar fragmentation at the NSCL. The measured half-life of this decay is $t_{1/2}=104\pm14$ ns. The energy and half-life of this $\gamma$ ray are in agreement with the previously reported results by \citet{Nishimura2014}, but the half-life measured in this work is slightly lower. We also determined an isomeric ratio of  $R=14\pm10\%$, which is much lower than the previously reported one obtained using a different reaction. This difference can be used to selectively produce either isomeric or ground-state beams of \textsuperscript{26}P in future experiments.  Finally, we have derived a semi-empirical constraint on the proton separation energy of \textsuperscript{26}P from the unobserved proton branch. A  measurement of the mass of \textsuperscript{26}P would give an experimental value for the proton separation energy, which would help to unambiguously determine whether this nucleus and its isomer are  bound or unbound to proton emission.

\begin{acknowledgments}
The authors gratefully acknowledge the contributions of the NSCL staff.
This work is supported by the U.S. National Science Foundation under Grants No. PHY-1102511, No. PHY-0822648, No. PHY-1350234, and No. PHY-1404442, the U.S. Department of Energy under Contract No. DE-FG02-97ER41020 and No. DE-SC0016052, the U.S. National Nuclear Security Agency under Grant No. DE-NA0000979 and No. DE-NA0002132; and the Natural Sciences and Engineering Research Council of Canada.

\end{acknowledgments}

\providecommand{\noopsort}[1]{}\providecommand{\singleletter}[1]{#1}%

\end{document}